\documentclass[sigconf]{acmart}

\AtBeginDocument{%
  }

\usepackage{algorithmic}
\usepackage{graphicx}
\usepackage{textcomp}
\usepackage{xcolor}

\usepackage{multirow}
\usepackage{booktabs}
\usepackage[ruled,linesnumbered]{algorithm2e}
\usepackage[normalem]{ulem}
\usepackage{balance}
\usepackage{threeparttable}
\usepackage{url}

\usepackage{graphicx}
\usepackage{textcomp}
\usepackage{xcolor}
\usepackage{hyperref}
\usepackage{subcaption}

\usepackage{enumitem}
\usepackage{dsfont}
\usepackage{booktabs}
\usepackage{makecell}
\usepackage{enumitem}

\usepackage{caption}
\copyrightyear{2026}
\acmYear{2026}
\setcopyright{cc}
\setcctype{by}
\acmConference[DAC '26]{63rd ACM/IEEE Design Automation Conference}{July 26--29, 2026}{Long Beach, CA, USA}
\acmBooktitle{63rd ACM/IEEE Design Automation Conference (DAC '26), July 26--29, 2026, Long Beach, CA, USA}
\acmDOI{10.1145/3770743.3803929}
\acmISBN{979-8-4007-2254-7/2026/07}

\begin{document}

% % \acmConference[]{}{}{}
% \acmConference[DAC '26]{63rd ACM/IEEE Design Automation Conference}{July 26--29, 2026}{Long Beach, CA, USA}

\title{Miter-Aware LUT Mapping: Aligning Structure and Solvability
for Efficient Logic Equivalence Checking}

% === Anonymous authors for submission ===
% \author{Anonymous Authors}

% \author{Jiaying Zhu}
% \email{jyzhu24@cse.cuhk.edu.hk}
% \affiliation{%
%   \institution{The Chinese University of Hong Kong}
% }

% \author{Zhengyuan Shi}
% \email{zyshi21@cse.cuhk.edu.hk}
% \affiliation{%
%   \institution{The Chinese University of Hong Kong}
% }

% \author{Mengxia Tao}
% \email{taomengxia@nctieda.com}
% \affiliation{%
%   \institution{National Center of Technology Innovation for EDA}
% }

% \author{Kezhi Li}
% \email{kzli24@cse.cuhk.edu.hk}
% \affiliation{%
%   \institution{The Chinese University of Hong Kong}
% }

% \author{Min Li}
% \email{min.li@seu.edu.cn}
% \affiliation{%
%   \institution{Southeast University}
% }

% \author{Qiang Xu}
% \email{qxu@cse.cuhk.edu.hk}
% \affiliation{%
%   \institution{The Chinese University of Hong Kong}
% }

\author{
Jiaying Zhu$^{1}$,
Zhengyuan Shi$^{1}$*,
Mengxia Tao$^{2}$,
Kezhi Li$^{1}$,
Min Li$^{3}$,
Qiang Xu$^{1}$*
}

% \authornote{Corresponding authors.}

\affiliation{
\institution{$^{1}$The Chinese University of Hong Kong, Hong Kong, China}
\city{}\country{}
}

\affiliation{
\institution{$^{2}$National Center of Technology Innovation for EDA, Nanjing, China}
\city{}\country{}
}

\affiliation{
\institution{$^{3}$Southeast University, Nanjing, China}
\city{}\country{}
}

\email{{jyzhu24, zyshi21, kzli24, qxu}@cse.cuhk.edu.hk, taomengxia@nctieda.com, min.li@seu.edu.cn}

\renewcommand{\shortauthors}{Jiaying Zhu et al.}

\begin{abstract}

%Logic Equivalence Checking (LEC) is fundamental to hardware verification but often suffers from synthesis-induced structural perturbations and XOR-intensive regions that significantly degrade the efficiency of SAT solvers. This paper presents a \emph{mapping-based modeling framework} that reformulates the miter before SAT solving. By constructing a \emph{LUT-based miter} rather than directly linking the two netlists at the bit level, our approach preserves structural correspondence and exposes logic relations. The framework combines three techniques: (1) \emph{equivalence-preserving LUT mapping} to align subgraphs between golden and implementation circuits, (2) \emph{Gaussian-guided XOR modeling mechanism} to simplify XOR-dense regions, and (3) \emph{solver-oriented LUT selection} guided by metrics of branching, propagation, and conflict interpretability. Experiments on the LECBench and MULT datasets show up to \textbf{12.7$\times$} speedup across multiple SAT solvers, demonstrating that structural modeling fundamentally enhances LEC efficiency by unifying LUT mapping and SAT reasoning into a solver-friendly reformulation.

Logic Equivalence Checking (LEC), a fundamental hardware verification task, is often bottlenecked by synthesis-induced structural perturbations and XOR-dense regions that degrade SAT solver performance. We contend that the \emph{modeling} of the miter is as critical as the SAT solver itself. To this end, we introduce a \emph{miter-aware mapping framework} that strategically formulates the problem before solving. By constructing a \emph{LUT-based miter}---instead of a traditional, flat netlist---our approach preserves critical structural correspondence between the two designs while making high-level logic relations explicit. Our framework uniquely integrates three techniques: \emph{equivalence-preserving mapping} to structurally align the two circuits, \emph{Gaussian-guided XOR modeling} to algebraically simplify dense arithmetic, and \emph{solver-oriented LUT selection} to generate a representation optimized for efficient SAT reasoning. Evaluated on comprehensive datasets, our method achieves up to a \textbf{92.1\%} reduction across state-of-the-art SAT solvers. This demonstrates that a solver-aware modeling paradigm, which unifies structural mapping with SAT reasoning, can fundamentally enhance LEC efficiency.

\end{abstract}
\maketitle

\setcounter{footnote}{0}
\begingroup
\renewcommand{\thefootnote}{*}
\footnotetext{Corresponding authors: Qiang Xu and Zhengyuan Shi.}
\endgroup

% \begin{IEEEkeywords}
% component, formatting, style, styling, insert.
% \end{IEEEkeywords}

\section{Introduction} \label{Sec:Intro}
Logic Equivalence Checking (LEC)~\cite{huang2012formal} proves that a golden design and its implementation compute identical Boolean functions. The \textit{de facto} LEC pipeline builds a \textit{miter} by connecting the outputs of two designs through XOR gates and checks whether the miter can ever evaluate to $1$. This turns equivalence checking into a single Boolean satisfiability (SAT) problem~\cite{SATHandbook}, and the approach has scaled thanks to modern SAT solvers~\cite{kissat,cadical,csat} and solver heuristics~\cite{een2005effective,hamadi2012control,liang2015understanding,shi2025dynamicsat}. 

However, two challenges limit the robustness of such pipeline on complex designs. First, logic synthesis (e.g. rewriting~\cite{mishchenko2006dag}, resubstitution~\cite{brayton2006scalable} and decomposition~\cite{machado2017boolean}) can significantly perturb logic cones and obscure matchable points between the miter’s two sides. The resulting lack of recognizable internal equivalences inhibits functional merging, a key simplification step shrinking the SAT instance within solvers~\cite{zhang2021deep, biere2024clausal}. Second, many real-world designs contain XOR-dense modules, such as arithmetic units and cryptographic units. Unfortunately, modern SAT solvers relying on the search-backtracking algorithm~\cite{marques2009conflict} poorly handle constraints derived from XOR chains, deducing a literal typically requires almost all inputs fixed and terminating searching branch after verbose decisions~\cite{chen2009building, soos2009extending, soos2019bird}. 

Although prior efforts attempt to simplify the SAT instances before solving~\cite{een2007applying, LogicOptimizationMeetsSAT, qian2025x}, operating solely on the flattened miters remains insufficient. 
We attribute that LEC efficiency is determined largely by \textit{how the miter is modeled before solving}. Rather than simply constructing a conventional miter and hoping the solver compensates, we advocate a modeling strategy that preserves structural correspondence and exposes strong implications prior to SAT. 

% \ST{Add figure 1: overview}
\begin{figure*}
    \centering
    \includegraphics[width=0.8\linewidth]{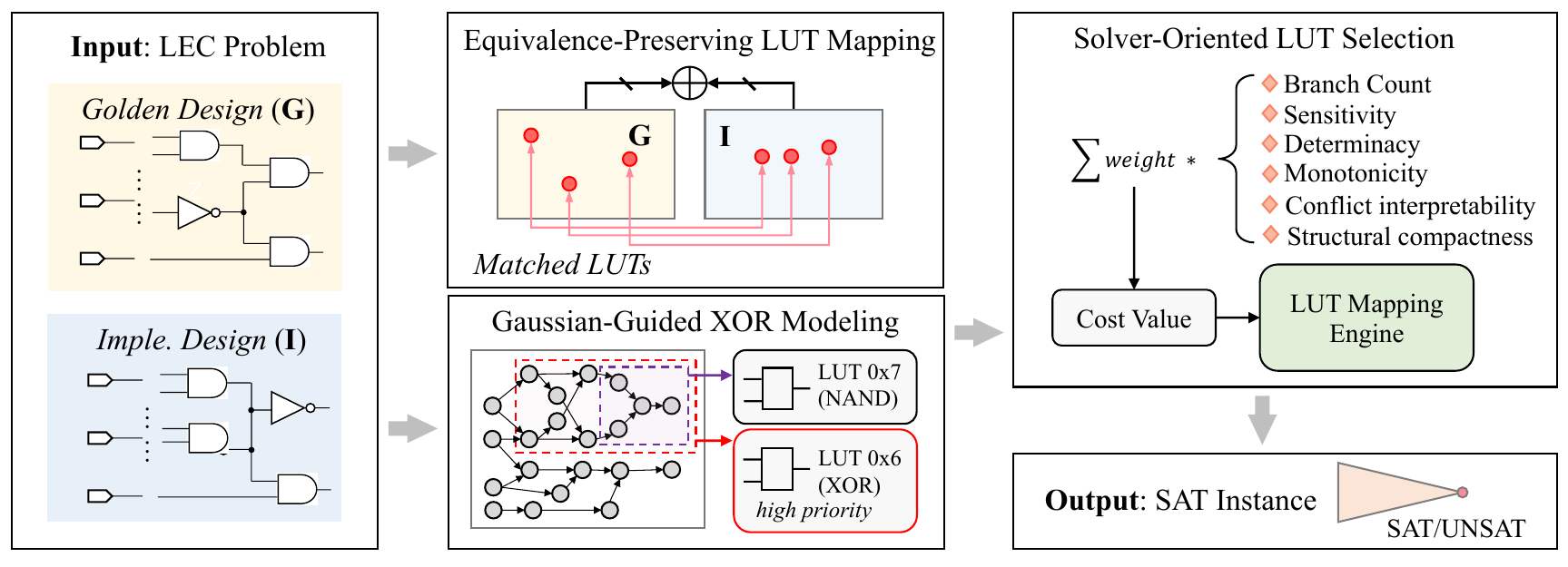}
    \caption{
    % Overview of the proposed mapping-based framework.
    % Given a miter circuit composed of two circuits under verification, 
    % our framework reformulates its representation through three integrated components:
    % (1) \emph{Equivalence-preserving LUT mapping}, which aligns structurally corresponding regions to preserve potential internal equivalences;
    % (2) \emph{Gaussian-guided XOR modeling}, which extracts affine relations within XOR-dense regions;
    % and (3) \emph{Solver-oriented LUT selection}, which optimizes each LUT according to solver-friendliness metrics 
    % The resulting LUT-based miter is structurally symmetric, algebraically simplified, 
    % and solver-efficient, benefiting both CNF- and circuit-based SAT solvers.
    Overview of the proposed mapping-based framework.
    Given a miter of two circuits under verification,
    our framework formulates its representation through three components:
    (1) \emph{Equivalence-preserving LUT mapping} aligning structurally corresponding regions to retain internal equivalences;
    (2) \emph{Gaussian-guided XOR modeling} extracting affine relations in XOR-dense regions; and
    (3) \emph{Solver-oriented LUT selection} optimizing each LUT by solver-friendliness metrics.
    % The resulting LUT-based miter is structurally symmetric, algebraically simplified, and solver-efficient for both CNF- and circuit-based solvers.
    }
    \label{fig:overview}
    % \vspace{-8pt}
\end{figure*}

In this work, we propose a novel mapping-based LEC formulation that constructs a simplified Look-up Table (LUT)-based miter (see Figure~\ref{fig:overview}) instead of directly linking the two netlists at the bit level. By mapping both designs to LUT abstraction, our approach preserves more internal identical cones, makes XOR structure explicit, and reduces solving complexity. We realize this agenda through three contributions: 

\begin{itemize}[noitemsep, topsep=0pt]
    \item An \textbf{equivalence-preserving LUT mapping algorithm} that actively aligns corresponding subgraphs across the two designs, forming more internal equivalent logic between implementation and golden designs.
    \item A \textbf{Gaussian-guided XOR modeling mechanism} that detects XOR-dense area, implements with the LUTs representing XOR logic, and leverages Gaussian elimination~\cite{Han2012WhenBS, lubke2024igmaxhs} to extract explicit affine relations.  
    \item A \textbf{solver-oriented LUT selection criterion} within the mapping algorithm that quantifies the solver-friendliness to minimize overall SAT solving difficulty.
\end{itemize}

We implement the proposed framework on top of a standard miter-based LEC flow and evaluate it using both structurally optimized and arithmetic-heavy benchmarks. 
%Experimental results demonstrate that our complete framework achieves an geometric mean average speedup of \textbf{7.0$\times$} compared with the baseline based on PAR2 metric. 
Experimental results demonstrate that our complete framework achieves up to 92.1\% reduction across solvers, with an 83.3\% average improvement over the baseline under the PAR2 metric.
Ablation results further show that the \emph{miter-aware mapping} alone reduces runtime by \textbf{62.5\%}, while the \emph{Gaussian-guided XOR modeling mechanism} alone yields a \textbf{36.8\%} reduction. 
Overall, our method can systematically reshape the problem representation—bridging LUT-mapping and SAT reasoning into a unified framework that delivers more compact CNFs and scalable equivalence checking performance.

%In summary, this work introduces a new LEC modeling paradigm that incorporates miter awareness, algebraic simplification, and solver-oriented mapping into a unified flow. The proposed framework not only improves practical verification efficiency but also provides a conceptual bridge between circuit optimization and SAT solving. 
% We expect this solver-friendly modeling strategy to inspire further cross-layer optimization for formal verification in advanced logic synthesis and EDA systems.

In summary, this work introduces a solver-aware modeling paradigm for LEC that unifies miter-aware mapping, algebraic simplification, and solver-oriented LUT selection. 
Rather than introducing a new mapping framework, we embed SAT/LEC-specific insights into standard LUT mapping so the resulting abstraction is solver-friendly rather than merely structurally compact.
This approach marks a significant departure from recent works~\cite{LogicOptimizationMeetsSAT, qian2025x} that apply LUT mapping to single-circuit SAT problems. While their methods optimize an individual netlist, our core contribution is the \textbf{joint mapping} of both golden and implementation circuits. This strategy is designed to explicitly preserve and rediscover the internal equivalences often obscured by synthesis---a challenge unique to equivalence checking. Consequently, the proposed framework not only improves practical verification efficiency but also forges a conceptual bridge between circuit optimization and SAT solving.

\section{Related Work} \label{Sec:Related}

% \ST{I merged [SAT and Circuit SAT] with [LEC]}

\subsection{Logic Equivalence Checking}
Logic Equivalence Checking (LEC) is a cornerstone of formal verification in EDA, aiming to ensure that two circuits, typically before and after synthesis, implement the same Boolean function~\cite{LECSurvey}. 
The standard approach is to build a \emph{miter circuit} by XORing corresponding outputs and ORing all XOR results together. Unsatisfiability of the resulting miter indicates functional equivalence. 

Typically, when two designs share a high degree of structural correspondence, equivalence points (i.e., functionally identical node pairs) can often be detected, making the miter instance of LEC easy to solve. However, aggressive logic synthesis disrupts the structural similarity~\cite{brayton2002multilevel, mishchenko2006dag, brayton2006scalable, machado2017boolean} but retain the hard-to-solve regions (i.e., XOR-dense regions)~\cite{chen2009building, soos2009extending, soos2019bird}. Recent efforts apply preprocessing to the flattened miter to simplify the SAT instance prior to solving~\cite{LogicOptimizationMeetsSAT,qian2025x}, yet overlooking the inherent cross-design correlations between the golden and implementation sides. Another line of research combines multiple engines, such as simulation-guided engine~\cite{SimSweeping, IncSATSweeping} or non-SAT alternatives~\cite{moon2004non}, to solve the miter instance. While effective on certain classes of designs, these methods still rely heavily on favorable problem features and initial decisions. In contrast to operating solely on a flattened miter, our work targets the initial \textbf{modeling} stage of LEC, reconstructing similar structure and neutralizing hard-to-solve regions before SAT solving.

\subsection{Look-up Table (LUT) Mapping}

% LUT-based representation has long been a cornerstone in FPGA synthesis~\cite{CongLUT}. 
% Recently, it has also attracted attention as a structure-preserving abstraction for SAT solving, where multiple logic gates are packed into solver-friendly LUTs to reduce CNF size and improve clause locality~\cite{LUTSAT2025}. 
% Each LUT encapsulates a small truth table, providing flexibility for customized encoding strategies and enabling control over structural granularity.

% Several works have shown that LUT-level abstraction can significantly improve SAT solver performance~\cite{LUTAbstractionSAT, LogicOptimizationMeetsSAT}. 
% For example, \textit{Logic Optimization Meets SAT}~\cite{LogicOptimizationMeetsSAT} explores LUT-based decomposition guided by branching complexity, demonstrating notable gains on arithmetic benchmarks. 
% However, existing methods focus mainly on single-circuit optimization and neglect miter-specific structures that arise in LEC problems. 
% Our work extends this direction by performing \textbf{miter-aware, solver-friendly LUT mapping}, jointly considering structural alignment, XOR handling, and CNF-level propagation properties.

Look-up Table (LUT) mapping is a fundamental step in logic synthesis, where a Boolean network is decomposed into a network of $k$-input LUTs for FPGA~\cite{ABC}. 
In a typical mapping procedure, the circuit is traversed in an output-to-input direction. 
For each node, multiple \textbf{candidate LUTs} are enumerated based on feasible input cuts and their corresponding Boolean functions, and the most suitable one is selected to optimize delay and latency.
Recently, it has also attracted attention as a structure-preserving abstraction for SAT solving, where multiple logic gates are packed into solver-friendly LUTs to reduce CNF size and improve clause locality~\cite{LogicOptimizationMeetsSAT,qian2025x}. 
Each LUT encapsulates a small truth table, offering flexibility for customized encoding strategies and fine-grained control over structural granularity.

The SAT-solving advantage of LUT-based abstraction arises from its ability to preserve local functional coherence \cite{qian2025x, biere2024clausal}. 
By grouping strongly related logic signals into a single LUT, the circuit representation becomes more compact and semantically cohesive, which helps remove redundant variables and clauses in the resulting CNF. 
Moreover, each LUT inherently defines a deterministic mapping between its inputs and output, enabling stronger inferences once partial assignments are known. 
This leads to more concentrated propagation, shorter implication chains, and overall improvement in solver efficiency.

However, existing approaches primarily target single-circuit optimization and overlook miter-specific structural patterns that arise in equivalence checking.
Our work extends this line of research by performing \textbf{miter-aware, solver-friendly LUT mapping}, jointly considering structural alignment, XOR reasoning, and CNF-level propagation properties.

\section{Methodology} \label{Sec:Method}

\subsection{Overview}
Given the golden design (\textbf{G}) and implementation design (\textit{Imple. Design}, \textbf{I}) in an LEC problem, our framework formulates it into an easy-to-solve SAT instance. As shown in Figure~\ref{fig:overview}, our framework targets three objectives: 
% Given a Logic Equivalence Checking (LEC) instance consisting of two circuits under formal verification, our goal is to perform \textbf{miter-aware, solver-friendly LUT mapping} that reformulates the problem representation and generates a CNF encoding more amenable to SAT solving. 
% % The proposed framework integrates structural alignment, algebraic simplification, and solver-oriented heuristics to jointly improve pre-solving reductions and in-solving propagation efficiency.

% \ST{Make sure the name of each component is coherent (in Introduction, Figure 1 and Methodology), e.g., equivalence-preserving LUT mapping algorithm, XOR elimination mechanism, solver-oriented LUT selection criterion}

% Specifically, it pursues three complementary objectives:
\begin{enumerate}
    \item \textbf{Preserve structural correspondence} across the two sides of the miter, facilitating equivalence identification and functionally reduction.
    \item \textbf{Gaussian-guided XOR modeling} to uncover the underlying affine relations among XOR variables and inject into the final CNF. 
    \item \textbf{Select solver-friendly LUTs} guided by multi-dimensional metrics that correlate with efficient branching, propagation, and conflict analysis.
\end{enumerate}

\subsection{Equivalence-preserving LUT Mapping}
\label{sec:miter}

\begin{figure}[t]
    \centering
    \includegraphics[width=0.9\linewidth]{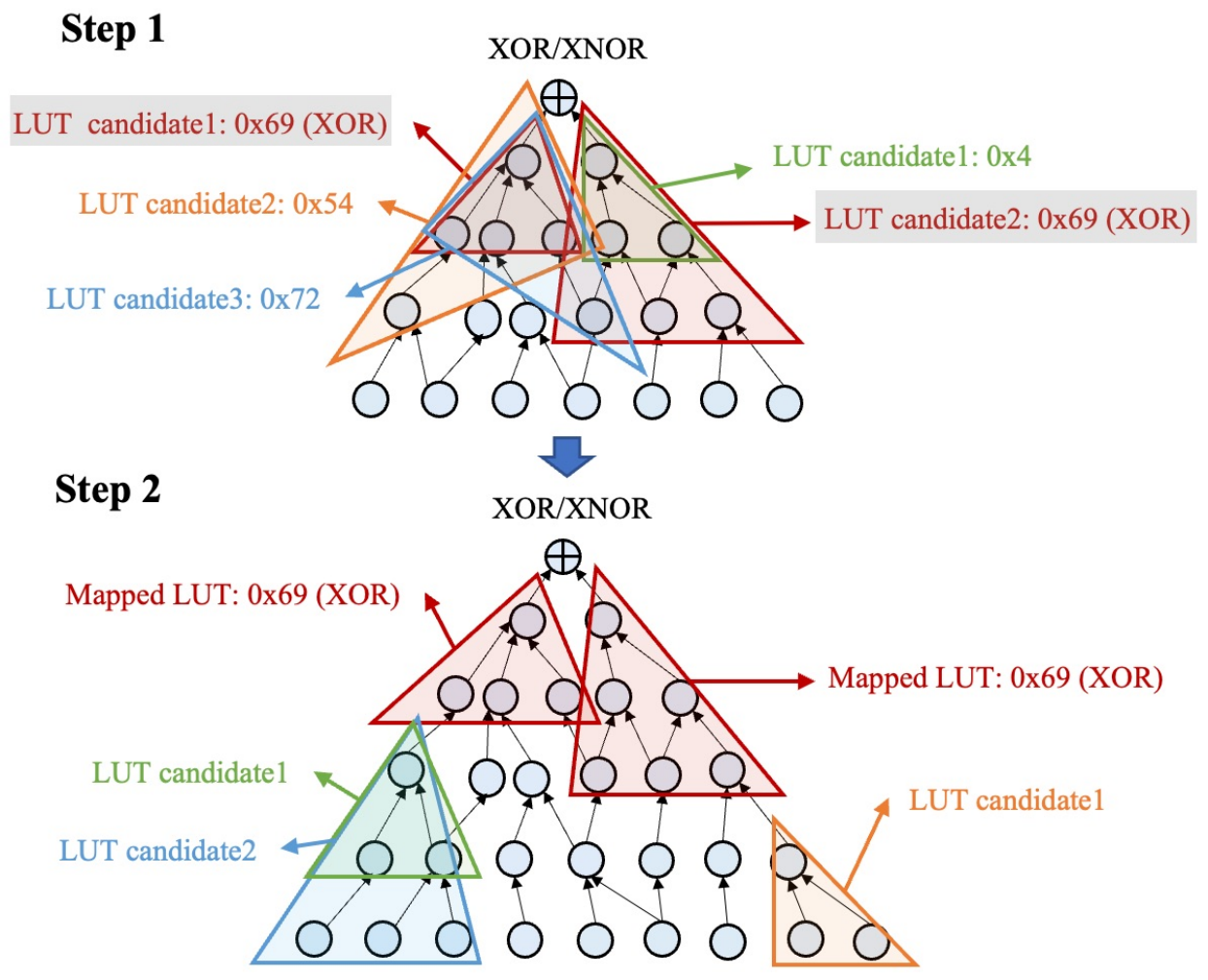}
    \caption{Illustration of equivalence-preserving mapping. 
    % From the primary outputs, XOR/XNOR gates are extracted to locate comparison points between the golden and implementation circuits. 
    % Mapping proceeds recursively in the output-to-input direction: multiple candidate LUTs are generated from feasible cuts on both sides, matched candidates are preferred to maintain structural correspondence, and unmatched ones are independently selected until reaching the primary inputs.
    }

    \label{fig:equ}
  
\end{figure}

% To mitigate the synthesis-induced structural perturbations, we introduce a \emph{miter-aware LUT mapping} scheme (Fig.~\ref{fig:equ}) that explicitly reconstructs structural correspondence between the two sides of the miter. We first identify XOR/XNOR structures near the primary outputs as natural comparison anchors between the golden and implementation circuits. The two fan-in nodes of these XOR gates serve as the respective boundaries of the two sides, providing an approximate yet practical alignment even after aggressive synthesis.
To mitigate the synthesis-induced structural perturbations, we introduce a \emph{miter-aware LUT mapping} scheme (Figure~\ref{fig:equ}) that explicitly reconstructs structural correspondence between the two sides of the miter. 
In a typical mapping procedure, the circuit is traversed in an output-to-input direction; for each node, multiple \textbf{candidate LUTs} are enumerated from feasible input cuts and their corresponding Boolean functions, and the most suitable one is selected to optimize delay.
Building upon this standard process, our mapping identifies XOR/XNOR structures near the primary outputs as natural comparison anchors between the golden and implementation circuits. 
The two fan-in nodes of these XOR gates serve as the respective boundaries of the two sides, providing an approximate yet practical alignment even after aggressive synthesis.

% From these boundary points, LUT decomposition proceeds recursively in a topological output-to-input direction. 
Starting from these boundary anchors, our mapper proceeds recursively to generate and\textbf{ match LUT candidates} on both sides. 
At each level, we create paired LUTs from nodes exhibiting structural similarity.
% —measured by topological depth and fan-in number\zjy{todo}—so that logically corresponding cones are encapsulated into aligned LUTs across the miter. 
This similarity is measured using a heuristic that combines topological depth, fan-in/fan-out counts, and the functional hash of the immediate logic cone.
Such matched mapping preserves functional equivalences that would otherwise be obscured by synthesis transformations.
% Unlike \textit{SAT-sweeping}~\cite{SimSweeping,IncSATSweeping}, our alignment is lightweight and avoids the cost of repeated SAT calls.
While \textit{SAT-sweeping}~\cite{SimSweeping,IncSATSweeping} detects equivalences via repeated SAT queries, our alignment provides a lightweight, orthogonal structural matching that complements such solver-based techniques without their computational cost.

It is worth noting that LUT mapping inherently \emph{absorbs} a group of primitive gates into each LUT, leaving only the LUT’s input and output variables exposed in the final CNF. Consequently, variables corresponding to internal gates disappear once absorbed. When mapping is guided to preserve equivalence points as LUT input or output boundaries, these equivalence variables remain explicit in the CNF and can be more effectively exploited by the solver’s redundancy elimination mechanisms~\cite{SATHandbook}. Moreover, equivalence points located closer to the primary outputs enable earlier clause subsumption and stronger pruning within the search space, amplifying solver efficiency.

If no direct correspondence can be found for a node pair—e.g., due to heavy resubstitution or decomposition—we independently select LUTs on each side according to the solver-friendliness metric introduced in Sec~\ref{sec:solverfriendly}, maintaining comparable structural granularity across designs. 
The resulting LUT network thus achieves improved symmetry, enhanced clause reuse, and stronger implication propagation, effectively bridging structural matching and solver-oriented optimization.

\subsection{Gaussian-guided XOR Modeling }
\label{sec:gaussian}

XOR-dominated circuits—such as arithmetic and cryptographic designs—pose major challenges to CDCL-based SAT solving, since their CNF encodings often require a large number of clauses and provide weak unit propagation. 

To this end, we first detect XOR relations within the AIG network and perform Gaussian elimination (GE) over the finite field GF (2) within XOR-dense regions.
%, not to remove variables, but to extract their underlying \emph{linear dependencies}..
Each XOR gate represents an equation among variables.
To construct equations with constant right-hand sides, we leverage simple transformations:
\begin{itemize}[noitemsep, topsep=0pt]
    \item $x_1 \oplus x_2 = x_3 \;\Rightarrow\; x_1 \oplus x_2 \oplus x_3 = 0$
    \item $x_1 \odot x_2 = x_3 \;\Rightarrow\; x_1 \oplus x_2 \oplus x_3 = 1$
\end{itemize}
where $\oplus$ denotes XOR, $\odot$ denotes XNOR.

For example, given the XOR relations:

\vspace{-18pt} 
\begin{align*}
x_1 \oplus x_2 \oplus x_3 \oplus x_5 &= 1,\\[-3pt]
x_1 \oplus x_2 \oplus x_4  &= 1,\\[-3pt]
x_2 \oplus x_3 \oplus x_5  &= 1,\\[-3pt]
x_1 \oplus x_4 \oplus x_5 &= 0,\\[-3pt]
\end{align*}

\vspace{-18pt}performing elementary row transformations yields:
% \vspace{-9pt}
\begin{align*}
x_1 \oplus x_2 \oplus x_5 &= 1,\\[-3pt]
x_1 \oplus x_3 &= 0,\\[-3pt]
x_2 \oplus x_4 &= 1,\\[-3pt]
x_4 \oplus x_5 &= 0. 
\end{align*}

Unlike conventional elimination that substitutes away variables, we selectively \textbf{retain XOR-related LUTs} whose variables participate in nontrivial affine relations discovered by Gaussian elimination. 
These LUTs are explicitly reconstructed as standalone XORs rather than merged into larger logic cones, ensuring that their participating variables remain visible and semantically linked in the final representation. 
Concretely, each affine relation derived from GE is directly instantiated as an XOR LUT connecting the involved variables, thereby materializing the linear dependency within the mapped circuit itself. 

For circuit-based solvers, this explicit construction enables early identification of XOR structures and reduces intermediate branching and propagation effort. 
For CNF-based solvers, the preserved XOR LUTs naturally translate into small XOR substructures during CNF conversion, allowing the solver to recognize and leverage these dependencies for stronger propagation and redundancy elimination ~\cite{biere2024clausal}. 
Overall, our approach integrates Gaussian-derived affine relations into the circuit as explicit XOR logic, ensuring their visibility across abstraction levels and improving both propagation and clause-level simplification.

\subsection{Solver-oriented LUT Selection}
\label{sec:solverfriendly}

For regions not dominated by XOR structures, we perform LUT implementations guided by a solver-friendliness metric.
% Each candidate LUT is evaluated along several dimensions, reflecting how its Boolean behavior affects the efficiency of SAT reasoning.

% Each LUT candidate $L$ is scored by:
% \[
% \text{Cost}(L) = \sum_i w_i f_i(L),
% \vspace{-6pt}
% \]
% where each $f_i(L)$ captures one solver-oriented property, and $w_i$ denotes its weight. 
% All $f_i(\cdot)$ are computed from the truth table of $L$, while the weights $\{w_i\}$ are automatically tuned via 
% Bayesian optimization~\cite{Snoek2012PracticalBO}, to minimize end-to-end SAT solving time on a held-out benchmark set.\zjy{todo}
% % which searches the parameter space to minimize the overall SAT solving time on a validation set of benchmark circuits.

% \subsubsection{ Favoring branching efficiency}
\vspace{4pt}
\noindent\textit{1) Branching efficiency:} 
% These metrics capture how fixing one variable influences other variables within the LUT.
These metrics capture how easily the solver can explore and prune the search space around $L$.

% \begin{enumerate}[label=(\alph*), leftmargin=2em]
\begin{enumerate}[label=(\alph*)]

\item \textbf{Branch count:} 
    % We measure the number of distinct input combinations producing different outputs. 
    % This metric quantifies how compactly a LUT partitions its input space into regions with deterministic outputs.
    We measure how compactly $L$ partitions the input space into regions with fixed output.
    Formally, a \emph{branch} is a pattern $p\in\{0,1,x\}^{|I|}$ (where $x$ denotes “don’t care”) such that every assignment covered by $p$ produces the same output.
    % Intuitively, fewer branches imply simpler early decisions and a search tree that is easier for SAT solvers to explore.
    % To make this metric scale-invariant across different input sizes, we normalize the branch count by the total number of input combinations:
    % \[
    % f_{\text{branch}}(L) = \frac{\mathrm{bc}(L)}{2^{|I|}},
    % \]
    % where a smaller value indicates a simpler, more solver-friendly function.
    Let $\mathrm{bc}(L)$ be the minimum number of such patterns needed to cover all satisfying assignments of $L$.
    We normalize by the input space size:
    \vspace{-4pt}\[
    f_{\text{branch}}(L)=\frac{\mathrm{bc}(L)}{2^{|I|}}.
    \]
    % so smaller values are preferred.
    Intuitively, a smaller branch count indicates that the function can be characterized by fewer consistent input regions, implying that decisions on input literals more rapidly constrain the search space.  
    From the SAT-solving perspective, such functions yield simpler decision trees, shorter BCP chains, and fewer case splits during clause learning, thus improving solver convergence.
    % Take 2-input AND and XOR as example, the AND function requires fewer branches than XOR ($3<4$), resulting in a lower normalized score. 
    % Hence, AND is considered more solver-friendly due to its simpler decision structure.
    For example, a 2-input AND can be covered by three branches (see Table~\ref{tab:and-branches}), while XOR requires four fully specified minterms, hence AND yields a lower $f_{\text{branch}}$ and is considered more solver-friendly.

    % \begin{table}[h]
    % \centering
    % \caption{Branch count of a 2-input AND}
    % \begin{tabular}{ccc}
    % \toprule
    % \textbf{input1} & \textbf{input2} & \textbf{output} \\
    % \midrule
    % 0 & $x$ & 0 \\
    % $x$ & 0 & 0 \\
    % 1 & 1 & 1 \\
    % \bottomrule
    % \end{tabular}
    % \label{tab:and-branches}
    % \end{table}

    % \begin{table}[h]
    % \centering
    % \caption{Branch count of a 2-input XOR}
    % \begin{tabular}{ccc}
    % \toprule
    % \textbf{input1} & \textbf{input2} & \textbf{output} \\
    % \midrule
    % 0 & 0 & 0 \\
    % 0 & 1 & 1 \\
    % 1 & 0 & 1 \\
    % 1 & 1 & 0 \\
    % \bottomrule
    % \end{tabular}
    % \label{tab:xor-branches}
    % \end{table}

    % \begin{table}[t]
    % \centering
    % \caption{Branch count comparison of 2-input gates.}
    % \vspace{-3pt} % 可选：控制标题与表格间距
    % \begin{minipage}[t]{0.46\linewidth}
    % \centering
    % \caption*{(a) 2-input AND}
    % \begin{tabular}{ccc}
    % \toprule
    % \textbf{input1} & \textbf{input2} & \textbf{output} \\
    % \midrule
    % 0 & $x$ & 0 \\
    % $x$ & 0 & 0 \\
    % 1 & 1 & 1 \\
    % \bottomrule
    % \end{tabular}
    % \label{tab:and-branches}
    % \end{minipage}
    % \hfill
    % \begin{minipage}[t]{0.46\linewidth}
    % \centering
    % \caption*{(b) 2-input XOR}
    % \begin{tabular}{ccc}
    % \toprule
    % \textbf{input1} & \textbf{input2} & \textbf{output} \\
    % \midrule
    % 0 & 0 & 0 \\
    % 0 & 1 & 1 \\
    % 1 & 0 & 1 \\
    % 1 & 1 & 0 \\
    % \bottomrule
    % \end{tabular}
    % \label{tab:xor-branches}
    % \end{minipage}
    % \vspace{-6pt}
    % \end{table}

\begin{table}[t]
\centering
\caption{Branch count comparison of 2-input gates.}
\vspace{-3pt}
\begin{minipage}[t]{0.44\linewidth} 
\centering
\small
\caption*{(a) 2-input AND}
\begin{tabular}{ccc}
\toprule
\textbf{input1} & \textbf{input2} & \textbf{output} \\
\midrule
0 & $x$ & 0 \\
$x$ & 0 & 0 \\
1 & 1 & 1 \\
\bottomrule
\end{tabular}
\label{tab:and-branches}
\end{minipage}
\hspace{0.05\linewidth} 
\begin{minipage}[t]{0.44\linewidth} 
\centering
\small
\caption*{(b) 2-input XOR}
\begin{tabular}{ccc}
\toprule
\textbf{input1} & \textbf{input2} & \textbf{output} \\
\midrule
0 & 0 & 0 \\
0 & 1 & 1 \\
1 & 0 & 1 \\
1 & 1 & 0 \\
\bottomrule
\end{tabular}
\label{tab:xor-branches}
\end{minipage}
\vspace{-5pt}
\end{table}

\item \textbf{Sensitivity:} 
    Sensitivity reflects how reactive $L$ is to single-bit flips. For each input assignment $a\in\{0,1\}^{|I|}$ and each input bit $i$, we flip bit $i$ to obtain $a\oplus e_i$ and check whether the output changes.
    Let $T_L(\cdot)$ be the Boolean function implemented by $L$.
    We define
    \vspace{-4pt}\[
    \mathrm{SensCount}(L)=
    \sum_{a\in\{0,1\}^{|I|}}\sum_{i=1}^{|I|} 
    1_{(T_L(a)\neq T_L(a\oplus e_i))},
    \]
    and normalize by the total number of flip trials:
    \[
    f_{\text{sensitivity}}(L)=
    \frac{\mathrm{SensCount}(L)}{|I|\cdot 2^{|I|}}.
    \]
    Lower $f_{\text{sensitivity}}$ indicates locally stable behavior and typically leads to a narrower effective branching space for CDCL-based solvers.
    % For each input variable $x_i$, we examine all possible input assignments and flip the $i$-th bit to check whether the output changes. 
    % Intuitively, a low sensitivity implies stable local behavior and thus easier propagation during solving.
    % % Conversely, a high sensitivity value means that flipping an input bit often changes the output, indicating a highly entangled input–output relation and a larger search space for SAT solvers.  
    % Formally, the \emph{sensitivity count} of a LUT L is defined as
    % % \[
    % % \mathrm{sens}(L) = \sum_{i=1}^{|I|} \bigl|\{\,a \in \{0,1\}^{|I|} \mid T_L(a) \neq T_L(a \oplus e_i)\,\}\bigr|,
    % % \]
    % \[
    % \mathrm{sens\_count}(L)
    % = \sum_{a\in\{0,1\}^{|I|}}\sum_{i=1}^{|I|}
    % 1_{\{\,T_L(a)\neq T_L(a\oplus e_i)\}},
    % \]
    % where $1_{\{\cdot\}}$ is an indicator function that equals 1 if the condition inside the brackets is true and 0 otherwise.
    % Here, $T_L(a)$ denotes the Boolean function implemented by LUT $L$, returning the output value of $L$ under the input assignment $a$.
    % $e_i$ denotes a unit vector that flips the $i$-th input bit.

    % To enable comparison across LUTs of different arities, we normalize it by the total
    % number of input–flip pairs:
    % \[
    % f_{\text{sensitivity}}(L) =
    % \frac{\mathrm{sens}(L)}{|I| \times 2^{|I|}},
    % \]
    % where smaller values correspond to more solver-friendly (less reactive) logic.

% \item \textbf{Entropy:} the expected reduction of output uncertainty after fixing one input variable.
    % \begin{align*}
    % H(O) &= -\sum_{v \in \{0,1\}} P(O=v)\log_2 P(O=v), \\
    % f_{\text{entropy}}(L) &= H(O).
    % \end{align*}

\end{enumerate}

% Combined:
% \[
% f_{\text{branch}}(L) = \lambda_1 f_{\text{branch-count}} + \lambda_2 f_{\text{sensitivity}} + \lambda_3 f_{\text{entropy}}.
% \]

% \subsubsection{ Favoring Boolean Constraint Propagation (BCP)}
\vspace{2pt}
\noindent\textit{2) Propagation strength.}
CDCL relies on Boolean Constraint Propagation (BCP)-partial assignments should force additional literals quickly.
% We assess propagation potential by:

\begin{enumerate}[label=(\alph*)]

\item \textbf{Determinacy:}
    Let $V=\{x_1,\dots,x_{|I|},O\}$ be LUT inputs and output.
    For each non-empty subset $S\subseteq V$, enumerate partial assignments $a_S$ on $S$ and test whether remaining variables $V\setminus S$ become uniquely determined.
    % (i.e., there is exactly one full assignment consistent with $a_S$ and $T_L$).
    Let $1_{\text{det}}(S,a_S)$ be 1 if so and 0 otherwise.
    We define
    \[
    \mathrm{DetCount}(L)=
    \sum_{\substack{S\subset V\\ S\neq\emptyset}}
    \sum_{a_S\in\{0,1\}^{|S|}}
    1_{\text{det}}(S,a_S),
    \]
   \[
    f_{\text{det}}(L)=
    \frac{\mathrm{DetCount}(L)}
    {\sum_{S\subset V,\,S\neq\emptyset} 2^{|S|}}.
    \]
    Higher $f_{\text{det}}$ means partial assignments tend to collapse the remaining degrees of freedom, enabling stronger BCP.
    % This metric measures how strongly the variables of a LUT depend on each other.
    % A LUT with high determinacy allows partial assignments to uniquely fix the remaining variables, reducing ambiguity in the Boolean space
    % and enabling stronger solver propagation.
    
    % Formally, let $V=\{x_1, x_2, \dots, x_{|I|}, O\}$ denote all input and output variables of LUT $L$.
    % For each non-empty subset $S \subset V$, we examine every partial assignment on $S$ and check whether the remaining variables $V\setminus S$ are uniquely determined.
    % Let $1(S,a)$ be an indicator equal to 1 if the partial assignment $a$ on $S$ uniquely fixes all variables in $V\setminus S$, and 0 otherwise.
    % The total number of deterministic cases is:
    % \[
    % \mathrm{DetCount}(L)
    % = \sum_{S\subset V,\,S\neq\emptyset}
    % \sum_{a\in\{0,1\}^{|S|}}
    % 1_{\text{det}}(S,a),
    % \]
    % and the normalized determinacy score is:
    % \[
    % f_{\text{determinacy}}(L)
    % = \frac{\mathrm{DetCount}(L)}
    % {\sum_{S\subset V,\,S\neq\emptyset}|S|},
    % \]
    % which reflects the proportion of partial assignments that uniquely determine all remaining variables.
    % A larger score indicates stronger variable dependence and higher solver-friendliness.

\item \textbf{Monotonicity:} 
    $T_L$ is monotone if for any $a,b\in\{0,1\}^{|I|}$ with $a_i\le b_i$ for all $i$, we have $T_L(a)\le T_L(b)$.
    Monotone functions typically translate into Horn-like CNFs with few negative literals, which in turn yield shorter propagation chains and more localized learned clauses, improving solver efficiency~\cite{Nishimura2004DetectingBS,Feng2024DRATPO}.
    Non-monotone functions (e.g., XOR) tend to generate balanced positive/negative clauses that weaken unit propagation.
    % % whether $a \le b \Rightarrow f(a) \le f(b)$ holds. 
    % % \[
    % % f_{\text{mono}}(L) = 1 - \frac{|\{(a,b)\mid a\le b, T_L(a)\le T_L(b)\}|}{2^{2|I|}}.
    % % \]
    % % Monotone LUTs yield Horn-like CNF encodings that support stronger local propagation.
    % Monotonicity describes the extent of a Boolean function preserves ordering with respect to its inputs.
    % A function $T_L$ is monotone if, for any two input assignments $a,b \in \{0,1\}^{|I|}$ satisfying $a \le b$ (i.e., $a_i \le b_i$ for all $i$), we have $T_L(a) \le T_L(b)$.
    
    % From the SAT-solving perspective, monotonic functions are advantageous because their CNF encodings tend to form \emph{Horn-like} clause structures with fewer negative literals~\cite{Nishimura2004DetectingBS,Feng2024DRATPO}.
    % Such CNFs induce shorter Boolean Constraint Propagation(BCP) paths and more localized learned clauses, thereby enhancing propagation efficiency and reducing solver overhead.
    % % In contrast, non-monotonic functions (e.g., XOR) generate balanced positive/negative clauses that significantly weaken unit propagation.
    
    % % We define the monotonicity score as the complement of the violation ratio:
    % % \[
    % % f_{\text{mono}}(L)
    % % = 1 - 
    % % \frac{
    % % |\{(a,b)\mid a\le b,\; T_L(a)\le T_L(b)\}|
    % % }{2^{2|I|}},
    % % \]
    % % where $T_L(a)$ denotes the Boolean function realized by LUT $L$.
    % % A smaller $f_{\text{mono}}(L)$ indicates fewer violations of monotonicity (i.e., stronger monotone behavior) and thus higher solver-friendliness.

\end{enumerate}

% \subsubsection{ Facilitating conflict analysis}
\vspace{2pt}
\noindent\textit{3) Conflict interpretability.}
We also favor LUTs 
% with locally explainable conflicts.
whose conflicts are easy to explain locally.
% whose conflicts are locally explainable.
For any non-empty $S\subseteq V$ and partial assignment $a_S$, we ask whether $a_S$ is \emph{infeasible}, i.e., no full assignment extending $a_S$ is consistent with $T_L$.
Let $1_{\text{conf}}(S,a_S)=1$ if $a_S$ is infeasible and $0$ otherwise.
We compute
\[
\mathrm{InterpCount}(L)=
\sum_{\substack{S\subset V\\ S\neq\emptyset}}
\sum_{a_S\in\{0,1\}^{|S|}}
1_{\text{conf}}(S,a_S),
\]
% \vspace{-2pt}
\[
f_{\text{interp}}(L)=
\frac{\mathrm{InterpCount}(L)}
{\sum_{S\subset V,\,S\neq\emptyset} 2^{|S|}}.
\]
A higher $f_{\text{interp}}$ means conflicts emerge from small local contexts, which leads to shorter learned clauses and faster convergence.

\vspace{2pt}
\noindent\textit{4) Structural compactness.}
Finally, we estimate how much internal logic a LUT absorbs from its fan-in cone, i.e., how many primitive logic gates (AIG nodes) are required to realize the Boolean function implemented by a $L$.
We approximate this count
% the number of primitive gates (AND/OR/NOT) required to realize the same Boolean function directly, 
by performing a sum-of-products–based minimization that derives prime implicants from the truth table~\cite{quine1952problem,mccluskey1956minimization}.
A larger gate count implies that the LUT encapsulates more local logic and thus eliminates more intermediate signals during mapping.
To ensure comparability across different input sizes, the score is normalized as
% \[
% f_{\text{compact}}(L)=\frac{\mathrm{GateCount}(L)}{|I|},
% \]
\[
f_{\text{compact}}(L)=\frac{|I|}{\mathrm{GateCount}(L)},
\]
where a smaller value indicates stronger logic absorption and fewer CNF variables introduced.

\vspace{2pt}
\noindent\textit{\textbf{Complexity and reusability.}}
% Although some metrics, such as $\text{DetCount}(L)$ and $\text{InterpCount}(L)$,
% require enumerating all non-empty variable subsets and all partial assignments, 
% the practical cost is negligible since our mapper only uses 2-, 3-, and 4-input LUTs. 
% For these small sizes, exhaustive evaluation is lightweight(under five seconds) and can be performed once offline. 
% The obtained scores are cached and reused in subsequent mappings, so no additional runtime overhead is introduced during the actual synthesis or SAT-solving flow.
Although metrics like $\text{DetCount}(L)$ and $\text{InterpCount}(L)$ require enumerating variable subsets and partial assignments, the cost is negligible since only 2/3/4-input LUTs are used. 
All scores are precomputed offline (in five seconds) and cached for reuse, introducing no overhead in synthesis or solving.

\vspace{2pt}
\noindent\textit{\textbf{Cost aggregation and weight learning.}}
The overall LUT cost combines each metric $f_i(L)$ with Bayesian-optimized~\cite{Bayesian2006} weight $w_i$, tuned on a small validation set to minimize solving time:
\[
\text{Cost}(L)=\sum_i w_i f_i(L).
\]
% where each $f_i$ encodes one solver-oriented property. 
% \vspace{-1pt}Analysis of the learned $\mathbf{w}$ shows distinct task preferences:
% in SAT-dominant cases, propagation-related determinacy $f_{\text{det}}$ gains higher weight, 
% facilitating faster implication convergence; 
% in UNSAT-dominant cases, branching metrics $f_{\text{branch}}$ and $f_{\text{sensitivity}}$ dominate, 
% as effective search partitioning accelerates conflict discovery.
While the weight $w_i$ is tuned on a subset, the results are robust across all validation sets. 
% Propagation ($f_{\textit{det}}$) and branching ($f_{\textit{branch}}$) metrics consistently received the highest weights, confirming the intuition that efficient BCP and search-space partitioning are the most critical factors for SAT performance. T
% he relative weighting between them(e.g., higher $f_{\textit{det}}$ for SAT, higher $f_{\textit{branch}}$ for UNSAT) aligns with known solver behaviors for satisfaction and proof-finding tasks, respectively.
Propagation ($f_{\textit{det}}$) and branching ($f_{\textit{branch}}$) metrics dominate, confirming that strong BCP and effective search partitioning are key to solver efficiency. 
Their relative emphasis (higher $f_{\textit{det}}$ for SAT, higher $f_{\textit{branch}}$ for UNSAT) aligns with the distinct behaviors of satisfaction and proof-oriented solvers.

% The overall LUT cost aggregates the metrics defined above. We tune the weight vector \(\mathbf{w}\) via Bayesian optimization on a held-out set to minimize average solving time.
% Each LUT candidate $L$ is scored by:
% \[
% \text{Cost}(L) = \sum_i w_i f_i(L),
% \vspace{-6pt}
% \]
% where each $f_i(L)$ captures one solver-oriented property, and $w_i$ denotes its weight. 

% Analyzing the learned \(\mathbf{w}\) reveals a task-dependent preference consistent with the metric intuition:
% on SAT-dominant subsets, the BCP-oriented determinacy \(f_{\text{det}}\) (Prop.~Strength) receives noticeably larger weight, indicating that quicker implication collapse accelerates reaching a satisfying assignment; on UNSAT-dominant subsets, branching-oriented \(f_{\text{branch}}\) and \(f_{\text{sensitivity}}\) are emphasized, as finer search-space partitioning and reactive flipping foster earlier contradictions and stronger conflict clauses.

\section{Experiments} \label{Sec:Experiment}

\subsection{Experiment Setting}
We construct a comprehensive benchmark suite to evaluate the efficiency of our proposed LEC formulation framework. This suite includes the circuit pairs derived from industrial LEC problem and the instances built from public sources: ForgeEDA~\cite{forgeeda2025}, ITC99~\cite{itc99}, EPFL~\cite{epfl}, and Opencore~\cite{opencore}. 
To build the equivalent circuit pairs, each circuit is synthesized and conjunctive with the original one to form a miter. To build the inequivalent circuit pairs, we introduce small functional perturbations by randomly modifying a few nodes in the synthesized AIGERs before miter construction. 

All circuits are synthesized with ABC~\cite{ABC}, and LUT networks are generated using a customized cost-driven mapper built on Mockturtle~\cite{soeken2018epfl}, restricted to 2-, 3-, and 4-input LUTs, since larger LUTs (i.e., $\ge$5 inputs) would produce exponentially more complex CNF encodings ($2^n$ clauses) and thus hinder subsequent SAT solving.
% to balance structural granularity and Circuit-to-CNF translation efficiency.
Each instance runs on a single Intel(R) Xeon(R) Platinum 8375C core with a 3{,}600 seconds timeout. %unsolved cases are marked as TIMEOUT and assigned 3600s for statistical analysis.
% Solver performance

% Performance is evaluated using the \textbf{PAR2} metric—commonly adopted in SAT competitions-where unsolved cases are penalized with twice the timeout value (i.e., 7200 s).

\subsection{Main results}
% \begin{table}[!t]
% \centering
% \caption{Average PAR2 comparison between Baseline and Ours. }
% \label{tab:main_results}
% \begin{tabular}{l|ccccc}
% \toprule
% \textbf{Solver} & Kissat & CaDiCaL & X-SAT & CSAT & CEC \\
% \midrule
% Baseline & 493.9 & 856.6 & 2287.3 & 4170.1 & \textbf{276.9} \\
% Ours     & \textbf{42.3}  & 67.4  & 289.4  & 868.7  & 92.8 \\ \midrule
% Reduction  & 91.4\% & \textbf{92.1\%} & 87.3\% & 79.2\% & 66.5\%  \\
% \bottomrule
% \end{tabular}
% \end{table}

\begin{table}[t]
\centering
\small
\caption{Solving performance comparison between Baseline and Ours across solvers.}
\label{tab:main_results}
\begin{tabular}{l l c c c c}
\toprule
\textbf{Solver} & \textbf{Setting} & \textbf{\#Solved} & \textbf{Impv.} & \textbf{PAR2} & \textbf{Red.} \\
\midrule
\multirow{2}{*}{Kissat} 
 & Baseline & 1,729 & -- & 493.9 & -- \\
 & Ours     & 1,731 & 0.12\% & 42.3 & 91.44\% \\
\midrule
\multirow{2}{*}{CaDiCaL} 
 & Baseline & 1,530 & -- & 856.6 & -- \\
 & Ours     & 1,723 & 12.61\% & 67.4 & \textbf{92.13\%} \\
\midrule
\multirow{2}{*}{X-SAT} 
 & Baseline & 1,125 & -- & 2287.3 & -- \\
 & Ours     & 1,498 & 33.16\% & 289.4 & 87.35\% \\
\midrule
\multirow{2}{*}{CSAT} 
 & Baseline & 373 & -- & 4170.1 & -- \\
 & Ours     & 996 & \textbf{167.02\%} & 868.7 & 79.17\% \\
\midrule
\multirow{2}{*}{CEC} 
 & Baseline & 1,735 & -- & 276.9 & -- \\
 & Ours     & 1,735 & 0.00\% & 92.8 & 66.49\% \\
\bottomrule
\end{tabular}
\vspace{-6pt}
\end{table}

To validate the effectiveness of our proposed framework, we compare the solving time of LEC instances formulated by our framework (Ours) and the baseline flow (Baseline). The performance is measured using average \textbf{PAR2} score, where any instance failing the 3{,}600s timeout is penalized with 7,200s. The PAR2 score for ``Ours'' includes both solving time and LUT mapping runtime, where the formulation overhead accounts for $\textless 10\%$. Besides, we also record the number of solved instance within 3,600s time limit as \textbf{\#Solved}. 

After formulating SAT instance of the given LEC problem, we employ two state-of-the-art CNF-based SAT solvers (Kissat~\cite{kissat} and CaDiCaL~\cite{cadical}), two circuit-based SAT solvers (X-SAT~\cite{qian2025x} and CSAT~\cite{csat}) and an advanced LEC engine (command \texttt{cec} in ABC~\cite{ABC}, denoted as CEC in Table~\ref{tab:main_results}), respectively. It should be noted that the LEC engine performs functional reduction (e.g., SAT-sweeping~\cite{SimSweeping}) and invokes CNF-based modern solver for final checking. Since CEC constructs the miter and translates it to CNF before calling solver, our approach is seamlessly inserted as a \emph{plugin-like pre-CNF optimization layer}. 

The results are summarized in Table~\ref{tab:main_results}, from which three key observations emerge. First, our framework consistently achieves significant positive gains across all solvers and engines. This is most dramatic with the CaDiCaL solver, where our approach reduces the PAR2 score by nearly an order of magnitude (from 856.6 to 67.4). Second, the improvement is particularly pronounced for the CNF-based SAT solvers, delivering reductions of 91.4\% for Kissat and 92.1\% for CaDiCaL. We attribute this to our framework's ability to leverage circuit-based methods for effective elimination—optimizations that CNF-based solvers cannot perform on their own. Third, our framework still provides a 66.49\% PAR2 reduction when used with the advanced CEC engine. Therefore, our approach is fully compatible with existing LEC tools, acting as a powerful pre-processing step that provides additional simplification, logic packing, and solver-friendly reformulation.

\subsection{Ablation Studies}

We further perform ablation studies to assess the effectiveness of the three key components in our framework using Kissat, the strongest solver among those tested.

% We further perform ablation studies to assess the effectiveness of the three key components in our framework.
% All experiments in this section are conducted with \textit{Kissat}, as it exhibits the strongest performance among the tested solvers.

% We further evaluate the contribution of each key component using Kissat, the strongest solver among those tested.

\subsubsection{Effectiveness of Equivalence-Preserving LUT Mapping} \label{Sec:Exp:Abs:EP}
% without miter-aware: Clausal Congruence Closure

To assess the impact of our \textbf{equivalence-preserving LUT mapping}, we conduct an ablation study comparing the configurations with (\textbf{Ours.}) or without (\textbf{Ours.w/o~Equiv}) the proposed method. Besides, we also evaluate the configurations with (\textbf{w~CCC}) or without (\textbf{w/o~CCC}) Clausal Congruence Closure (CCC)~\cite{biere2024clausal}. CCC is a powerful CNF-level circuit simplification technique, which detects AND, XOR, and ITE gate structures directly from CNF formulas and performs congruence closure to merge equivalent literals. We include such comparison to demonstrate our LUT mapping provides benefits that are complementary to the subsequent CNF-based optimizations. 

Therefore, we have four settings in this experiments: 
\begin{itemize}[noitemsep, topsep=0pt]
    \item \textbf{Ours.w~CCC}: our proposed equivalence-preserving mapping algorithm combined with the redundancy elimination in \textit{Clausal Congruence Closure} (denoted as CCC).
    \item \textbf{Ours.w/o~CCC}: our mapping method without CCC.
                    %(i.e., circuit-level only).
    \item \textbf{Ours.w/o~Equiv~w~CCC}: the variant without our equivalence-preserving mapping but still using CCC.
    \item \textbf{Ours.w/o~Equiv~w/o~CCC}: the variant that disables the equivalence-preserving mapping and also removing CCC.
\end{itemize}

% \begin{figure}[t]
%     \centering
%     \includegraphics[width=0.8\linewidth]{Figure/ablation-equ.png}
%     \caption{
%     Solving performance under the equivalence-preserving LUT mapping ablation. 
%     % CCC = Clausal Congruence Closure.
%     % \textit{Ours.w~CCC} combines both circuit-level equivalence-preserving mapping and CNF-level redundancy elimination, achieving the best overall performance. 
%     }
%     \label{fig:ablation-equ}
% \end{figure}

Here, Clausal Congruence Closure serves as a \textit{CNF-level circuit simplification technique}, which detects AND, XOR, and ITE gate structures directly from CNF formulas and performs congruence closure to merge equivalent literals, thereby reducing structural redundancy and improving SAT solving efficiency. 
In contrast, our equivalence-preserving mapping operates at the \textit{circuit level} during LUT decomposition, maintaining structural correspondence between the golden and implementation sides of the miter. 
We include both \emph{with} and \emph{without} CCC settings to demonstrate that our method is \emph{complementary} to CNF-level simplification—our mapping enhances structural regularity before CNF translation, 
while CCC further eliminates redundancy after translation.

Table ~\ref{tab:ablation-equ} presents the cumulative solving performance for all four variants on the entire dataset.
Comparing \textbf{Ours.w/o~Equiv~w/o~CCC} and \textbf{Ours.w/o~Equiv~w~CCC} shows that CCC alone contributes a noticeable performance gain by simplifying redundant CNF structures.
However, when our equivalence-preserving mapping is enabled (\textbf{Ours.w/o~CCC}), an even larger improvement is observed, indicating that preserving equivalence alignment early at the circuit level leads to substantially better clause reuse and stronger implication propagation during solving.

\begin{table}[!t]
\centering
\small
\caption{Solving performance comparison under the equivalence-preserving LUT mapping ablation.}
\label{tab:ablation-equ}
\begin{tabular}{lcccc}
\toprule
\textbf{Setting} & \textbf{\#Solved} & \textbf{Impv.} & \textbf{PAR2} & \textbf{Red.} \\
\midrule
Ours.w/o Equiv w/o CCC & 1,713 & --        & 396.5 & -- \\
Ours.w/o Equiv w CCC   & \textbf{1,733} & \textbf{1.17\%} & 314.5 & 20.67\% \\
Ours.w/o CCC         & 1,726 & 0.76\%    & 56.2  & 85.83\% \\
Ours.w CCC           & 1,731 & 1.05\%    & \textbf{42.3} & \textbf{89.33\%} \\
\bottomrule
\end{tabular}
\vspace{-4pt}
\end{table}

\subsubsection{Gaussian-guided XOR Modeling}
% without gauss: Clausal Congruence Closure

% To evaluate the contribution of the proposed \textbf{Gaussian-guided XOR modeling}, we perform an ablation study on the \textit{MULT} dataset, which contains 20 highly XOR-dense multiplier circuits. 
To assess the contribution of the proposed \textbf{Gaussian-guided XOR modeling}, we compare the solver performance with (\textbf{Ours.}) and without (\textbf{Ours.w/o Gauss}) this component. While we observe an average 5.3\% PAR2 reduction across the entire dataset by using Gaussian-guided XOR modeling, the gain is especially significant on XOR-dense datapath circuits. 

Therefore, we perform a deeper analysis on 20 circuits constructed from \textit{32$\times$32 multipliers} implemented with diverse architectures~\cite{Abbasi2015FPGA,CEKLI2024155435}. 
These multipliers pose one of the most challenging verification tasks~\cite{Kaufmann2023improving} due to their deep and XOR-dense structures. To maintain a consistent comparison and demonstrate complementarity with CNF-level simplifications as Section~\ref{Sec:Exp:Abs:EP}, we again compare the configurations with (\textbf{w CCC}) or without CCC (\textbf{w/o CCC}). 

% The XOR elimination operates before CNF encoding, identifying linear dependencies among XOR gates and simplifying them through Gaussian elimination. In contrast, CCC~\cite{biere2024clausal} functions at the CNF level by merging syntactically equivalent literals. As in the previous study, these two techniques target different stages of the flow and can therefore be combined seamlessly.

\begin{figure}
    \captionsetup{skip=4pt} 
    \centering
    \includegraphics[width=0.7\linewidth]{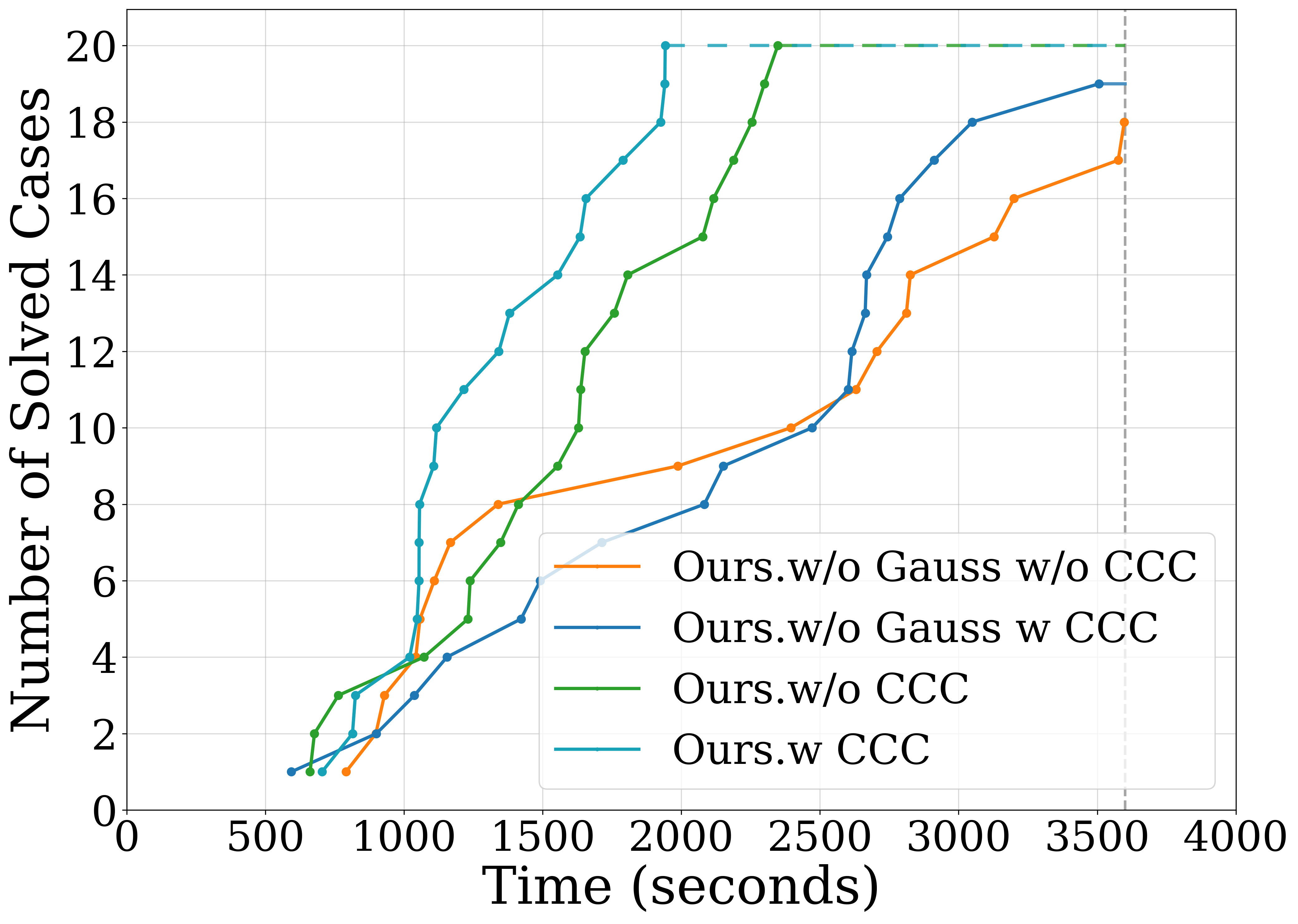}
    \caption{
    Solving performance on 20 multipliers under the Gaussian-guided XOR modeling ablation.
    % CCC = Clausal Congruence Closure.
    % The proposed Gaussian simplification and CCC operate at different levels of the flow and jointly achieve the highest overall efficiency. 
    }
    \label{fig:ablation-gauss}
    \vspace{-10pt}
    
\end{figure}

The results are shown in Figure~\ref{fig:ablation-gauss}. First, a large improvement on solving efficiency is seen when our method is applied (\textbf{Ours.w/o Gauss w CCC} v.s. \textbf{Ours. w CCC}). With our method, all 20 XOR-dense instances are solved within 2{,}000 seconds. In contrast, the configuration without our Gaussian modeling solves only 7 instances by the 2,000s mark and 19 instances by 3,600s. Second, the two techniques are not conflicting, as they operate at different stages of the flow and can be applied simultaneously—\textbf{Ours.w~CCC} achieves the best overall performance, yielding an average speedup of \textbf{36.8\%} over \textbf{Ours.w/o~Gauss~w~CCC}.

\subsubsection{Solver-Oriented LUT Selection Criterion}
% without metric: 25dac, 07sat, (xsat)

To evaluate the effectiveness of our proposed solver-friendliness metric, we isolate it from the other two components and activate only the metric-guided LUT selection module (denoted as \textbf{Ours.metric}). Two baselines are included for comparison:
\begin{itemize}[noitemsep, topsep=0pt]
    \item \textbf{Comp.1}, from prior work~\cite{LogicOptimizationMeetsSAT}, which employs a branching-based LUT cost function emphasizing structural complexity.
    \item \textbf{Comp.2}, from prior work~\cite{een2007applying}, which uses area-oriented cost functions based on CNF clause counts during mapping.
\end{itemize}

\begin{figure}
    \captionsetup{skip=4pt} 
    \centering
    \includegraphics[width=0.7\linewidth]{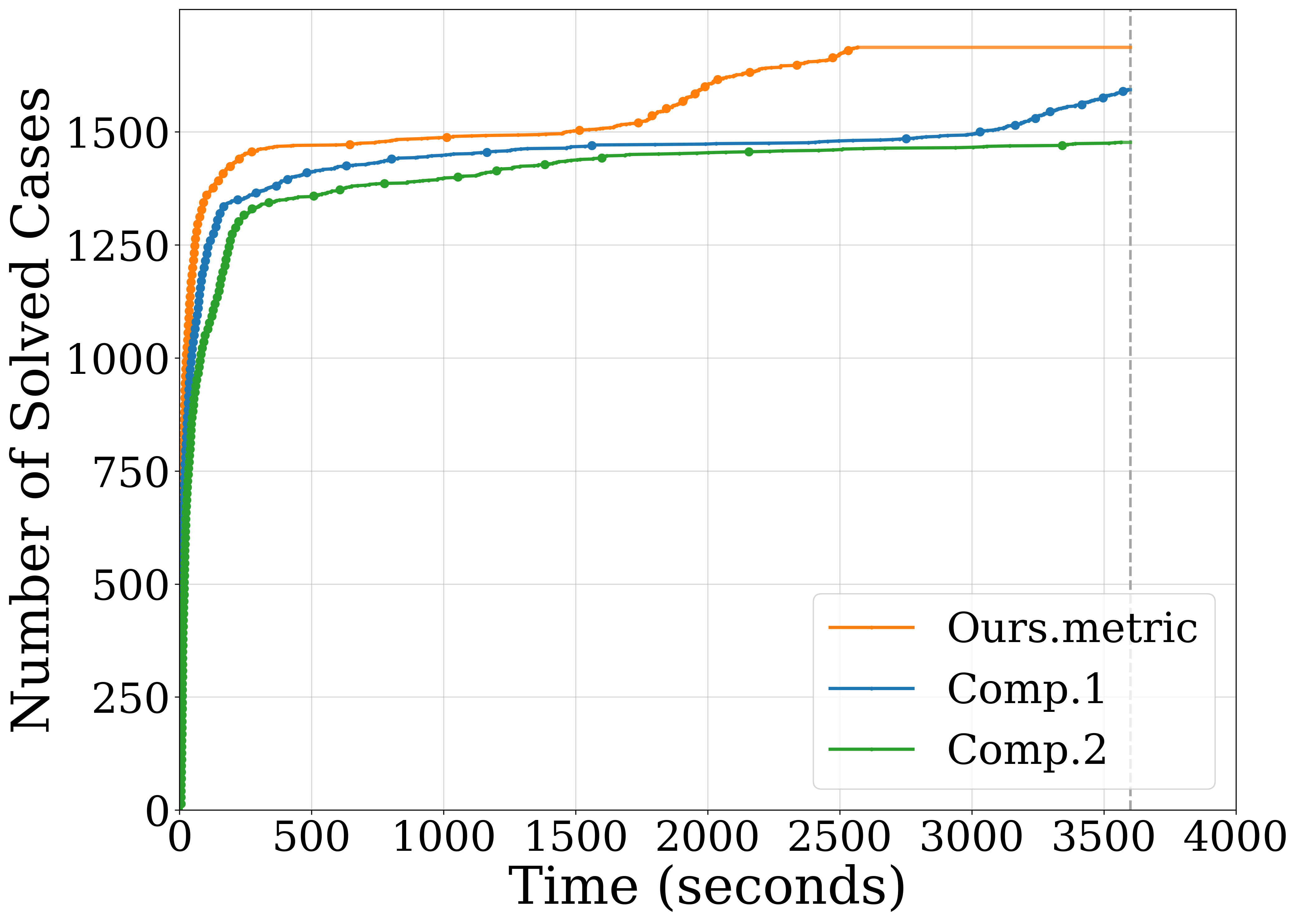}
    \caption{
    Solving performance under the solver-oriented LUT selection ablation.
    % The x-axis represents runtime (seconds), and the y-axis denotes the number of solved instances. 
    }
    \label{fig:abltaion_metric}
    \vspace{-4pt}   
\end{figure}

As shown in Figure~\ref{fig:abltaion_metric}, \textbf{Ours.metric} consistently outperforms both \textbf{Comp.1} and \textbf{Comp.2} across the runtime range. Compared with \textbf{Comp.1}, it solves \textbf{5.5\%} more instances within 3{,}600 seconds and yields an overall \textbf{34.1\%} average speedup. These results demonstrate that our metric effectively captures solver-oriented behaviors beyond structural complexity, leading to a more solver-efficient formulation.

\section{Conclusion}
\label{Sec:Conclusion}

% This paper introduced a mapping-based modeling framework that models the representation of Logic Equivalence Checking problems. By integrating equivalence-preserving LUT mapping, Gaussian-guided XOR modeling, and solver-oriented LUT selection into a unified flow, our method reformulates the verification problem into a structurally symmetric, algebraically simplified, and solver-efficient form. Extensive experiments on LECBench and MULT benchmarks validate the effectiveness and generality of our approach, consistently improving solving efficiency across multiple solver paradigms. 
% This work demonstrates that LEC performance can be substantially improved by modeling reformulation. We believe this study bridges the long-standing gap between logic synthesis and SAT-based formal verification.

This paper presented a mapping-based modeling framework for Logic Equivalence Checking. 
By integrating equivalence-preserving LUT mapping, Gaussian-guided XOR modeling, and solver-oriented LUT selection into a unified flow, our method formulates the miter into a structurally symmetric, algebraically simplified, and solver-efficient form. 
Experiments on comprehensive benchmarks confirm substantial and consistent improvements across multiple solver paradigms, demonstrating that modeling formulation can markedly enhance LEC efficiency and bridge the gap between logic synthesis and SAT-based formal verification.

\balance

\section*{Acknowledgments}
This work was supported in part by the Hong Kong Research Grants Council (RGC) under Grant No. N\_CUHK451/25, 14202824, C6003-24Y, and T46-415/25-R.

\newpage

\balance
\bibliographystyle{ACM-Reference-Format}
\bibliography{reference}

@inproceedings{zhang2021deep,
  title={Deep integration of circuit simulator and SAT solver},
  author={Zhang, He-Teng and Jiang, Jie-Hong R and Amar{\'u}, Luca and Mishchenko, Alan and Brayton, Robert},
  booktitle={2021 58th ACM/IEEE Design Automation Conference (DAC)},
  pages={877--882},
  year={2021},
  organization={IEEE}
}

@book{huang2012formal,
  title={Formal equivalence checking and design debugging},
  author={Huang, Shi-Yu and Cheng, Kwang-Ting Tim},
  volume={12},
  year={2012},
  publisher={Springer Science \& Business Media}
}

@inproceedings{biere2024clausal,
  title={Clausal congruence closure},
  author={Biere, Armin and Fazekas, Katalin and Fleury, Mathias and Froleyks, Nils},
  booktitle={27th International Conference on Theory and Applications of Satisfiability Testing (SAT 2024)},
  pages={6--1},
  year={2024},
  organization={Schloss Dagstuhl--Leibniz-Zentrum f{\"u}r Informatik}
}

@inproceedings{qian2025x,
  title={X-SAT: An Efficient Circuit-Based SAT Solver},
  author={Qian, Yuhang and Chen, Zhihan and Zhang, Xindi and Cai, Shaowei},
  booktitle={2025 62nd ACM/IEEE Design Automation Conference (DAC)},
  pages={1--7},
  year={2025},
  organization={IEEE}
}

@article{LECSurvey,
  title={Combinational and sequential equivalence checking: Theory and practice},
  author={Kuehlmann, Andreas and Krohm, Florian},
  journal={IEEE Transactions on Computer-Aided Design of Integrated Circuits and Systems},
  volume={19},
  number={12},
  pages={1428--1449},
  year={2000},
  publisher={IEEE}
}

@book{SATHandbook,
  title        = {Handbook of Satisfiability},
  author       = {Biere, Armin and Heule, Marijn J. H. and van Maaren, Hans and Walsh, Toby},
  year         = {2021},
  publisher    = {IOS Press},
  series       = {Frontiers in Artificial Intelligence and Applications},
  volume       = {336}
}

@inproceedings{ABC,
  title        = {ABC: A System for Sequential Synthesis and Verification},
  author       = {Brayton, Robert K. and Mishchenko, Alan},
  booktitle    = {Proceedings of the 22nd International Conference on Computer Aided Verification (CAV)},
  year         = {2010},
  pages        = {24--40}
}

@inproceedings{LogicOptimizationMeetsSAT,
  title        = {Logic Optimization Meets SAT: A Novel Framework for Circuit-SAT Solving},
  author       = {Shi, Zhengyuan and Tang, Tiebing and Zhu, Jiaying and Khan, Sadaf and Zhen, Hui-Ling and Yuan, Minjie and Chu, Zhufei and Xu, Qiang},
  booktitle    = {Proceedings of the 62nd ACM/IEEE Design Automation Conference (DAC)},
  year         = {2025},
  pages        = {1--6}
}

@inproceedings{SimSweeping,
  title={Simulation-based boolean equivalence checking using {AIGs}},
  author={Mishchenko, Alan and Chatterjee, Satrajit and Brayton, Robert},
  booktitle={Proceedings of the International Workshop on Logic and Synthesis (IWLS)},
  pages={7--12},
  year={2006}
}

@inproceedings{liang2015understanding,
  title={Understanding VSIDS branching heuristics in conflict-driven clause-learning SAT solvers},
  author={Liang, Jia Hui and Ganesh, Vijay and Zulkoski, Ed and Zaman, Atulan and Czarnecki, Krzysztof},
  booktitle={Haifa Verification Conference},
  pages={225--241},
  year={2015},
  organization={Springer}
}

@inproceedings{shi2025dynamicsat,
  title={Dynamicsat: Dynamic configuration tuning for sat solving},
  author={Shi, Zhengyuan and Jiang, Wentao and Zhang, Xindi and Luo, Jin and Liang, Yun and Chu, Zhufei and Xu, Qiang},
  booktitle={31st International Conference on Principles and Practice of Constraint Programming (CP 2025)},
  pages={34--1},
  year={2025},
  organization={Schloss Dagstuhl--Leibniz-Zentrum f{\"u}r Informatik}
}

@inproceedings{een2005effective,
  title={Effective preprocessing in SAT through variable and clause elimination},
  author={E{\'e}n, Niklas and Biere, Armin},
  booktitle={International conference on theory and applications of satisfiability testing},
  pages={61--75},
  year={2005},
  organization={Springer}
}

@incollection{hamadi2012control,
  title={Control-based clause sharing in parallel SAT solving},
  author={Hamadi, Youssef and Jabbour, Said and Sais, Jabbour},
  booktitle={Autonomous Search},
  pages={245--267},
  year={2012},
  publisher={Springer}
}

@inproceedings{IncSATSweeping,
  title={Incremental SAT sweeping},
  author={Een, Niklas and Mishchenko, Alan},
  booktitle={Proceedings of the IEEE/ACM International Conference on Computer-Aided Design (ICCAD)},
  pages={223--230},
  year={2007},
  organization={IEEE}
}

@article{soeken2018epfl,
  title={The EPFL logic synthesis libraries},
  author={Soeken, Mathias and Riener, Heinz and Haaswijk, Winston and Testa, Eleonora and Schmitt, Bruno and Meuli, Giulia and Mozafari, Fereshte and Lee, Siang-Yun and Calvino, Alessandro Tempia and Marakkalage, Dewmini Sudara and others},
  journal={arXiv preprint arXiv:1805.05121},
  year={2018}
}

@inproceedings{moon2004non,
  title={Non-miter-based combinational equivalence checking by comparing BDDs with different variable orders},
  author={Moon, In-Ho and Pixley, Carl},
  booktitle={International Conference on Formal Methods in Computer-Aided Design},
  pages={144--158},
  year={2004},
  organization={Springer}
}

@article{forgeeda2025,
  author    = {Shi, Zhengyuan and Li, Zeju and Ma, Chengyu and Zhou, Yunhao and Zheng, Ziyang and Liu, Jiawei and Pan, Hongyang and Zhou, Lingfeng and Li, Kezhi and Zhu, Jiaying and Yan, Lingwei and He, Zhiqiang and Xue, Chenhao and Jiang, Wentao and Yang, Fan and Sun, Guangyu and Yang, Xiaoyan and Chen, Gang and Shi, Chuan and Chu, Zhufei and Yang, Jun and Xu, Qiang},
  title     = {ForgeEDA: Towards Verifiable and Customizable Circuit Benchmarks},
  journal   = {arXiv preprint arXiv:2505.02016},
  year      = {2025},
  url       = {https://arxiv.org/abs/2505.02016}
}

@INPROCEEDINGS{itc99,
  author={Davidson, S.},
  booktitle={International Test Conference 1999. Proceedings (IEEE Cat. No.99CH37034)}, 
  title={ITC'99 Benchmark Circuits - Preliminary Results}, 
  year={1999},
  volume={},
  number={},
  pages={1125-1125},
  doi={10.1109/TEST.1999.805857}}

@INPROCEEDINGS{epfl,
  author={Luca Amaru and Pierre-Emmanuel Gaillardon and Eleonora Testa and Giovanni De Micheli},
  title={The EPFL Combinational Benchmark Suite}, 
  booktitle={24th International Workshop on Logic \& Synthesis (IWLS)}, 
  year={2019},
  volume={},
  number={},
  doi={10.1109/TEST.1999.805857}}

@article{opencore,
  title={OpenCore},
  author={OpenCore, Team},
  year={1999},
  journal={https://opencores.org/}
}

@article{kissat,
  author    = {Armin, Biere and Tobias, Faller and Katalin, Fazekas and Mathias, Fleury and Nils, Froleyks and Florian, Pollitt},
  title     = {CaDiCaL, Gimsatul, IsaSAT and Kissat entering the SAT Competition 2024},
  journal   = {Proc. of SAT Competition},
  year      = {2024},
  pages     = {8-10}
}

@inproceedings{cadical,
author = {Biere, Armin and Faller, Tobias and Fazekas, Katalin and Fleury, Mathias and Froleyks, Nils and Pollitt, Florian},
title = {CaDiCaL 2.0},
year = {2024},
isbn = {978-3-031-65626-2},
publisher = {Springer-Verlag},
address = {Berlin, Heidelberg},
url = {https://doi.org/10.1007/978-3-031-65627-9_7},
doi = {10.1007/978-3-031-65627-9_7},
pages = {133–152},
numpages = {20},
location = {Montreal, QC, Canada}
}

@inproceedings{csat,
  title={A circuit-based SAT solver for logic synthesis},
  author={Zhang, He-Teng and Jiang, Jie-Hong R and Mishchenko, Alan},
  booktitle={2021 IEEE/ACM International Conference On Computer Aided Design (ICCAD)},
  pages={1--6},
  year={2021},
  organization={IEEE}
}

@inproceedings{mishchenko2006dag,
  title={DAG-aware AIG rewriting a fresh look at combinational logic synthesis},
  author={Mishchenko, Alan and Chatterjee, Satrajit and Brayton, Robert},
  booktitle={Proceedings of the 43rd annual Design Automation Conference},
  pages={532--535},
  year={2006}
}

@inproceedings{brayton2006scalable,
  title={Scalable logic synthesis using a simple circuit structure},
  author={Alan Mishchenko and Robert K. Brayton},
  booktitle={Proc. IWLS},
  volume={6},
  pages={15--22},
  year={2006}
}

@inproceedings{machado2017boolean,
  title={Boolean decomposition for aig optimization},
  author={Machado, Lucas and Cortadella, Jordi},
  booktitle={Proceedings of the Great Lakes Symposium on VLSI 2017},
  pages={143--148},
  year={2017}
}

@article{marques2009conflict,
  title={Conflict-driven clause learning SAT solvers},
  author={Marques-Silva, Joao and Lynce, In{\^e}s and Malik, Sharad},
  journal={Handbook of satisfiability},
  pages={131--153},
  year={2009},
  publisher={ios Press}
}

@inproceedings{een2007applying,
  title={Applying logic synthesis for speeding up SAT},
  author={E{\'e}n, Niklas and Mishchenko, Alan and S{\"o}rensson, Niklas},
  booktitle={International Conference on Theory and Applications of Satisfiability Testing},
  pages={272--286},
  year={2007},
  organization={Springer}
}

@article{brayton2002multilevel,
  title={Multilevel logic synthesis},
  author={Brayton, Robert K and Hachtel, Gary D and Sangiovanni-Vincentelli, Alberto L},
  journal={Proceedings of the IEEE},
  volume={78},
  number={2},
  pages={264--300},
  year={2002},
  publisher={IEEE}
}

@inproceedings{chen2009building,
  title={Building a hybrid SAT solver via conflict-driven, look-ahead and XOR reasoning techniques},
  author={Chen, Jingchao},
  booktitle={International Conference on Theory and Applications of Satisfiability Testing},
  pages={298--311},
  year={2009},
  organization={Springer}
}

@inproceedings{soos2009extending,
  title={Extending SAT solvers to cryptographic problems},
  author={Soos, Mate and Nohl, Karsten and Castelluccia, Claude},
  booktitle={International Conference on Theory and Applications of Satisfiability Testing},
  pages={244--257},
  year={2009},
  organization={Springer}
}

@inproceedings{soos2019bird,
  title={BIRD: engineering an efficient CNF-XOR SAT solver and its applications to approximate model counting},
  author={Soos, Mate and Meel, Kuldeep S},
  booktitle={Proceedings of the AAAI Conference on Artificial Intelligence},
  volume={33},
  number={01},
  pages={1592--1599},
  year={2019}
}

@inproceedings{Han2012WhenBS,
  title={When Boolean Satisfiability Meets Gaussian Elimination in a Simplex Way},
  author={Cheng-Shen Han and Jie-Hong Roland Jiang},
  booktitle={International Conference on Computer Aided Verification},
  year={2012},
  url={https://api.semanticscholar.org/CorpusID:18689957}
}

@article{lubke2024igmaxhs,
  title={IGMaxHS--An Incremental MaxSAT Solver with Support for XOR Clauses},
  author={L{\"u}bke, Ole},
  journal={arXiv preprint arXiv:2410.15897},
  year={2024}
}

@inproceedings{Nishimura2004DetectingBS,
  title={Detecting Backdoor Sets with Respect to Horn and Binary Clauses},
  author={Naomi Nishimura and Prabhakar Ragde and Stefan Szeider},
  booktitle={International Conference on Theory and Applications of Satisfiability Testing},
  year={2004},
  url={https://api.semanticscholar.org/CorpusID:15232413}
}

@article{Feng2024DRATPO,
  title={DRAT Proofs of Unsatisfiability for SAT Modulo Monotonic Theories},
  author={Nick Feng and Alan J. Hu and Sam Bayless and Syed M. Iqbal and Patrick Trentin and Michael W. Whalen and Lee Pike and John Backes},
  journal={ArXiv},
  year={2024},
  volume={abs/2401.10703},
  url={https://api.semanticscholar.org/CorpusID:267061064}
}

@article{Abbasi2015FPGA,
author = {Abbasi, Shuja and Zulhelmi, Zulhelmi and Alamoud, Abdulrahman},
year = {2015},
month = {07},
pages = {},
title = {FPGA Design, Simulation and Prototyping of a High Speed 32-bit Pipeline Multiplier Based on Vedic Mathematics},
volume = {12},
journal = {IEICE Electronics Express},
doi = {10.1587/elex.12.20150450}
}

@article{CEKLI2024155435,
title = {A high speed pipelined radix-16 Booth multiplier architecture for FPGA implementation},
journal = {AEU - International Journal of Electronics and Communications},
volume = {185},
pages = {155435},
year = {2024},
issn = {1434-8411},
doi = {https://doi.org/10.1016/j.aeue.2024.155435},
url = {https://www.sciencedirect.com/science/article/pii/S1434841124003212},
author = {Serap Cekli and Ali Akman},
}

@article{Kaufmann2023improving,
title = {Improving AMulet2 for verifying multiplier circuits using SAT solving and computer algebra},
author = {Kaufmann, Daniela and Biere, Armin},
year = {2023},
month = {01},
pages = {1-12},
volume = {25},
journal = {International Journal on Software Tools for Technology Transfer},
doi = {10.1007/s10009-022-00688-6}
}

@article{quine1952problem,
  author    = {Quine, Willard V.},
  title     = {The problem of simplifying truth functions},
  journal   = {The American Mathematical Monthly},
  volume    = {59},
  number    = {8},
  pages     = {521--531},
  year      = {1952}
}

@article{mccluskey1956minimization,
  author    = {McCluskey, Edward J.},
  title     = {Minimization of Boolean functions},
  journal   = {Bell System Technical Journal},
  volume    = {35},
  number    = {6},
  pages     = {1417--1444},
  year      = {1956}
}

@inbook{Bayesian2006,
author = {Mockus, Jonas},
year = {2006},
month = {01},
pages = {473-481},
title = {Bayesian Approach to Global Optimization},
volume = {37},
isbn = {978-94-010-6898-7},
doi = {10.1007/BFb0006170}
}

\end{document}